\documentclass[]{spie}  

\usepackage{amsmath,amsfonts,amssymb}
\usepackage{subcaption}
\usepackage{graphicx}
\usepackage[colorlinks=true, allcolors=blue]{hyperref}

\title{GPI 2.0: Optical Designs for the Upgrade of the Gemini Planet Imager Coronagraphic system
}

\author[a]{Meiji M. Nguyen}
\author[a]{Bryony F. Nickson}
\author[a]{Emiel H. Por}
\author[a]{R{\'e}mi Soummer}
\author[b, c]{John G. Hagopian}
\author[d]{Bruce Macintosh}
\author[e]{Jeffrey Chilcote}
\author[a]{Laurent Pueyo}
\author[a]{Marshall Perrin}
\author[f]{Quinn Konopacky}
\affil[a]{Space Telescope Science Institute, 3700 San Martin Dr, Baltimore, MD 21218, USA}
\affil[b]{NASA Goddard Space Flight Center, 8800 Greenbelt Rd, Greenbelt, MD 20771, USA}
\affil[c]{Lambda Consulting LLC/Advanced NanoPhotonics, 4437 Windsor Farm Rd Harwood, MD 20776, USA}
\affil[d]{Stanford University, 450 Serra Mall, Stanford, CA 94305, USA}
\affil[e]{University of Notre Dame, 1220 E. Angela Blvd.
South Bend, IN 46617, USA}
\affil[f]{University of California, San Diego, 5998 Alcala Park, San Diego, CA 92110, USA}
\authorinfo{Further author information: (Send correspondence to Meiji M. Nguyen: E-mail: mnguyen@stsci.edu)}

\begin{document} 
\maketitle

\begin{abstract}
The Gemini Planet Imager (GPI) is an integral field spectrograph (IFS) and coronagraph that is one of the few current generation instruments optimized for high-contrast direct imaging of substellar companions. The instrument is in the process of being upgraded and moved from its current mount on the Gemini South Observatory in Cerro Pachón, Chile, to its twin observatory, Gemini North, on Mauna Kea (a process colloquially dubbed “GPI 2.0”).  We present the designs that have been developed for the part of GPI 2.0 that pertains to upgrading various optical components of the GPI coronagraphic system. More specifically, we present new designs for the apodizer and Lyot stop (LS) that achieve better raw contrast at the inner working angle of the dark zone as well as improved core throughput while retaining a similar level of robustness to LS misalignment. To generate these upgraded designs, we use our own publicly available software package called APLC-Optimization that combines a commercial linear solver (Gurobi) with a high contrast imaging simulation package (HCIPy) in order to iteratively propagate light through a simulated model of an apodized phase lyot coronagraph (APLC), optimizing for the best coronagraph performance metrics. The designs have recently finished being lithographically printed by a commercial manufacturer and will be ready for use when GPI 2.0 goes on-sky in 2023. 
\end{abstract}


\section{INTRODUCTION}
\label{sec:intro}  
The Gemini Planet Imager \cite{2014PNAS..11112661M} is a near-infrared high-contrast direct imager designed to search for and characterize sub-stellar companions \cite{2015Sci...350...64M} and disks \cite{2020AJ....160...24E} at small angular resolution relative to their host star. The primary mission of the instrument is an 890 hr observation campaign, the Gemini Planet Imager Exoplanet Survey (GPIES)\cite{2019AJ....158...13N}. After nearly 7 years of continuous operation since 2013, the instrument was decommissioned in August 2020 in order to be moved from the Gemini South telescope in Cerro Pachón, Chile to the Gemini North telescope atop Mauna Kea in Hawaii. Since both Gemini North and South are twin observatories, the instrument would sit in the same position when installed on Gemini North (behind the Cassegrain focus of the 8.1-m  telescope) as it was used in on Gemini South.  

\begin{figure} [ht]
\begin{center}
\begin{tabular}{c} 
\includegraphics[scale=.5]{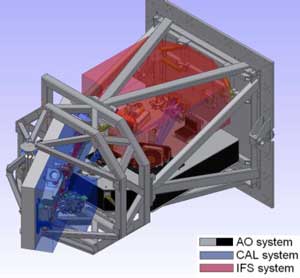}
\hspace{1cm}
\includegraphics[scale=.6]{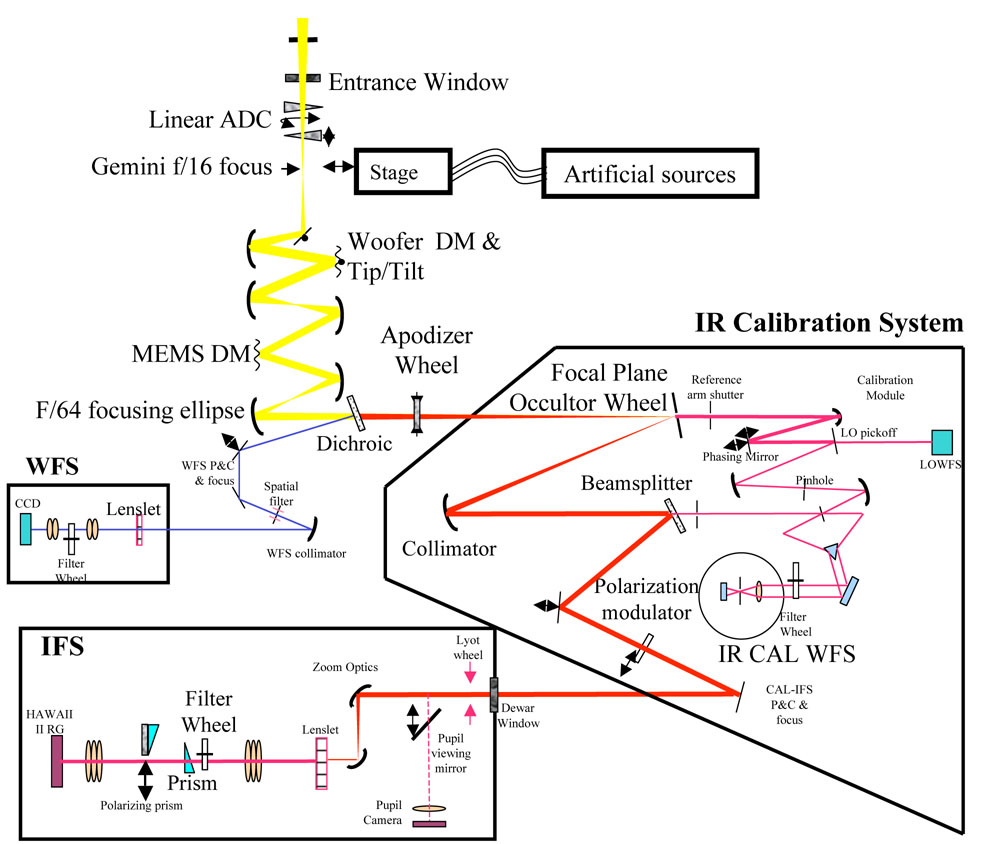}
\end{tabular}
\end{center}
\caption[ap] 
{ \label{fig:gpi_layout} 
(Left) A 3D solid model of GPI showing the three main components of the instrument: the AO system (black), the Calibration system (blue), and the IFS system (red). (Right) A more detailed schematic diagram showing all the various sub-components of GPI. In this study, we are only concerned with upgrades to the coronagraph subsystem, specifically changes to the apodizer wheel and Lyot wheel. Both figures are taken from De Rosa et al. (2020) \cite{2020JATIS...6a5006D}.}
\end{figure}

There are several major components of the planned upgrade of GPI including an upgraded AO-system which uses a faster realtime computer and a pyramid wavefront sensor, better observatory software, a more sensitive IR tip/tilt camera, various improvements to the IFS, and updated coronagraphic masks (see Fig. 4 of Chilcote et al. 2020\cite{2020SPIE11447E..1SC}). In this study, we will be presenting the results for the updated coronagraphic masks, specifically the new apodizers and Lyot stops which will go into the apodizer and Lyot wheels of GPI respectively (see Fig. \ref{fig:gpi_layout}). We first start by presenting the theoretical framework and software tools we use to generate the apodizer and LS designs in Section \ref{sec:opt}. Then we discuss the three sets of new designs (where a "set" is an apodizer with its corresponding LS) that have been settled upon and physically sent out for manufacture in Section \ref{sec:three}. Finally, we examine the theoretical performance of the GPI 2.0 designs and the expected improvement they should provide using various high contrast imaging metrics such as the throughput, raw contrast, and robustness to misalignment tolerance in Section \ref{sec:perf}. Though the new apodizers and Lyot stops have just recently finished fabrication, the move of GPI from Gemini South was delayed due to the pandemic and other factors, so laboratory testing of these optics was not yet available at the time of this report.

\section{THE OPTIMIZATION PROCESS}
\label{sec:opt}

\begin{figure*} [ht] 
\begin{center}
\includegraphics[scale=.5]{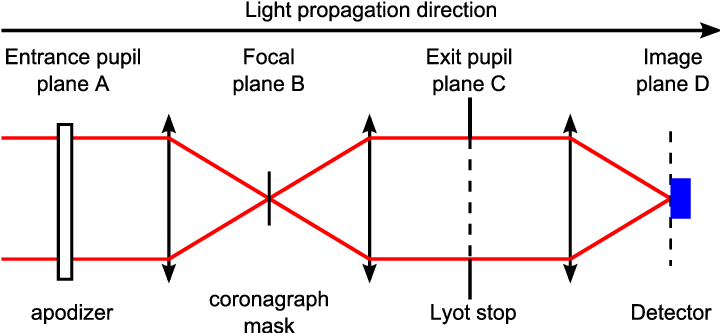}
\end{center}
\caption
{ \label{fig:aplc_schematic} 
A basic schematic of the planes of propagation in an APLC. This figure is taken from Fig. 1 of N'diaye et al. (2015)\cite{2015ApJ...799..225N}.}
\end{figure*}

\begin{figure*}
\begin{center}
\vspace{5mm}
\hspace*{-3.5cm} 
\includegraphics[scale=1.3]{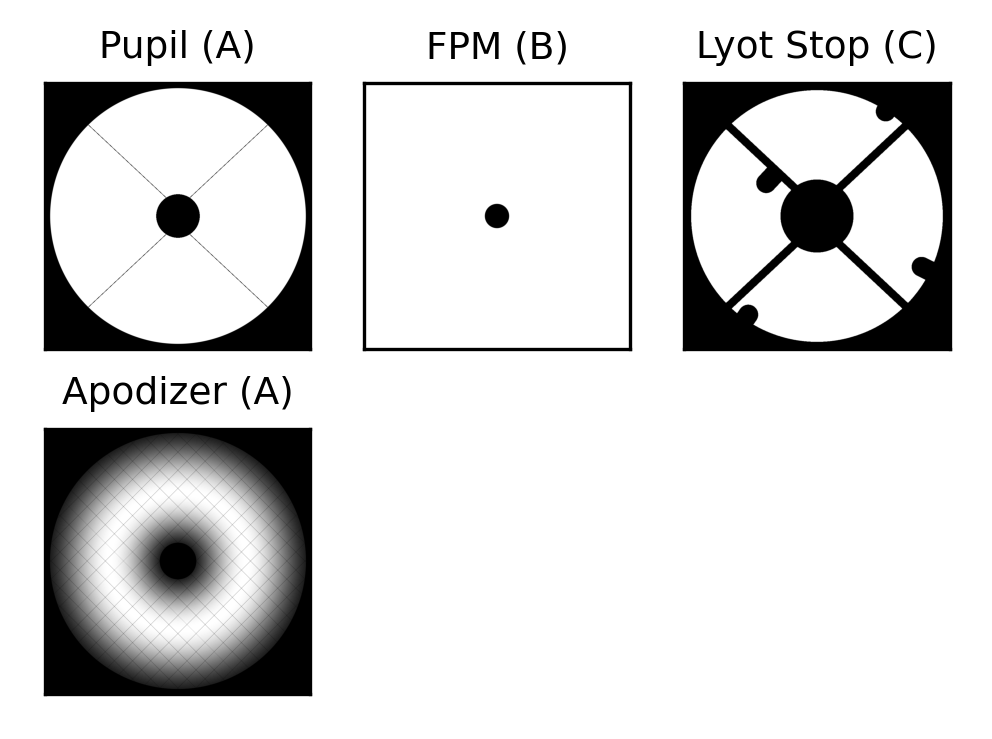}
\end{center}
\caption
{ \label{fig:aplc_masks} 
Example images of the pupil/apodizer, focal plane, and Lyot stop masks in the appropriate order to match the direction of light propagation as the diagram above. These are the binary representations of some of the previous generation GPI 1.0 masks. See Section \ref{sec:three} for the actual GPI 2.0 masks.}
\vspace{5mm}
\end{figure*} 

The infrastructure we use to generate the optical designs is all contained within a publicly available STScI github repository called APLC-Optimization\cite{2022SPIEbry, por2020aplc} \footnote{https://github.com/spacetelescope/aplc\_optimization}.
There are two main components of this optimization code: (1) An APLC simulator built using the high-contrast imaging python package HCIPy \cite{2018SPIE10703E..42P}; and (2) A commercial linear solver called Gurobi \footnote{https://www.gurobi.com/products/gurobi-optimizer/} that iteratively propagates light through the APLC simulator, varying the shape of the apodizer and minimizing the loss function of the output coronagraphic point spread function (PSF) with respect to various performance metrics such as the raw contrast or core throughput. To simulate an APLC, we need to have four optical planes of propagation (see Fig. \ref{fig:aplc_schematic}). A detailed mathematical description of the light propagation through each plane in an APLC is beyond the scope of this paper (see Soummer et al. 2007 \cite{2007OExpr..1515935S} for a full treatment), but the two basic operations are: (1) Fourier (or inverse Fourier) transforms to propagate between pupil planes and focal planes (or vice versa), and (2) multiplications (or convolutions depending on whether the multiplication is done in a conjugate Fourier space) by masks to selectively filter light in the appropriate planes (namely multiplication/convolution by either an apodizer mask, a focal plane mask, or a Lyot stop mask). We can follow the light propagation through the APLC using the following basic algorithm (represented in Fig. \ref{fig:aplc_schematic}):

\textbf{Plane A)} The light comes in from the entrance pupil and gets multiplied by the apodizer mask.

\textbf{Plane B)} We take the Fourier transform of this light to propagate it to the focal plane, then multiply the result by the focal plane mask (FPM).

\textbf{Plane C)} We take the inverse Fourier transform of this light to propagate it to the exit pupil plane, then multiply the result by the Lyot stop mask. 

\textbf{Plane D)} We take the Fourier transform of this light to propagate it to the final image plane where the detector sits.

\section{THREE NEW DESIGNS}
\label{sec:three}

\subsection{Apodizers}
\label{sec:ap}

The optimization process assumes a fixed primary pupil shape and fixed focal plane mask since these surfaces are not mutable on GPI. This leaves the apodizer and Lyot stop planes (in planes A and C respectively) as the two surfaces we can design for. After various rounds of optimizations and testing, a total of three separate apodizer designs were settled upon (with each apodizer design paired with a corresponding Lyot stop optimized for it). For simplicity, we will refer to these three designs as \textbf{LS03Symm}, \textbf{DualPlane}, and \textbf{DualPlaneSymm}, with the reasoning behind the names to be described below. For all three of the designs, we have overlaid an evenly spaced diffraction grating (also known as an astrometric grid) on top. The astrometric grid creates a diffraction pattern of regularly spaced out copies of the core PSF, called "satellite spots" (or "satspots" for short), that are useful for astrometry and calibration purposes.

\textbf{Design 1: LS03Symm)} The first apodizer design, LS03Symm, assumed a fixed Lyot stop geometry during the optimization process based on a horizontally symmetrized version of one of the previous generation GPI 1.0 Lyot stops (specifically mask "080m12\_03", hence the reason why we have given it the name "LS03" along with the suffix "-Symm", see Table 2 in \ref{footnote:gpi}). The reason we used a symmetrized version of the LS03 Lyot stop was because of memory constraints on the server computers on which the optimizations ran. We found that there was a memory bottleneck during the optimization process and introducing symmetry into the LS reduced the computational overhead enough to bypass this bottleneck. We optimized this design for the GPI 1.0 K1 FPM (m/2 = 3.476 $\lambda/D$) which is designed for H band, but we expect this mask to perform similarly in other spectral bands as long as the ratios in wavelengths are equal. We set the target contrast for the dark zone to be  $10^{-8}$ over a 20$\%$ bandpass,  corresponding to one order of magnitude higher contrast than the specification of the GPI 1.0 coronagraphs\cite{Soummer09, Soummer_2011}. 

\textbf{Design 2: DualPlane)} The second apodizer design, DualPlane, allowed both the apodizer and Lyot stop to vary during the optimization process (hence the name "Dual"Plane). This design was optimized to produce a "grayscale" solution, meaning the surface of the apodizer can vary in its transmissivity anywhere from 0 (fully absorptive) to 1 (fully transmissive). This is in contrast to the other two apodizer designs which are "binary" solutions, meaning the surface is manufactured to be either fully absorptive or fully transmissive with no partial transmissivity in between. For the details on how this particular design was developed, see Por et al. (2022) \cite{2022SPIEemiel}.

\textbf{Design 3: DualPlaneSymm)} The third apodizer design, DualPlaneSymm, is partially based on Design 2, assuming a fixed Lyot stop geometry during the optimization process that is a vertically symmetrized version of Design 2's LS (hence the name DualPlane+"-Symm"). We symmetrized the LS for the same memory constraint reason as Design 1's symmetrized LS. We also used similar optimization parameters (target contrast, bandpass, K1 FPM mask) that Design 1 used.

\begin{figure} [p!]
\begin{center}
\begin{tabular}{c|c|c} 
\textbf{LS03Symm Apod} & \textbf{DualPlane Apod} & \textbf{DualPlaneSymm Apod}\\

\includegraphics[height=5cm]{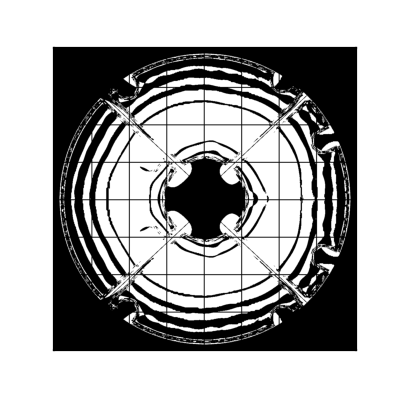} &
\includegraphics[height=5cm]{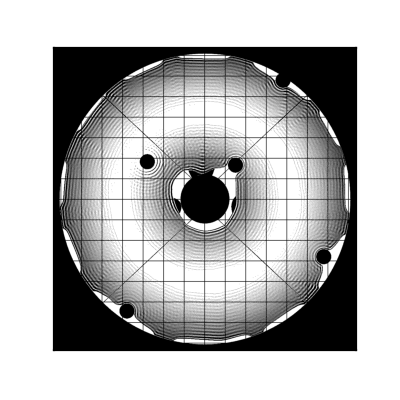} &
\includegraphics[height=5cm]{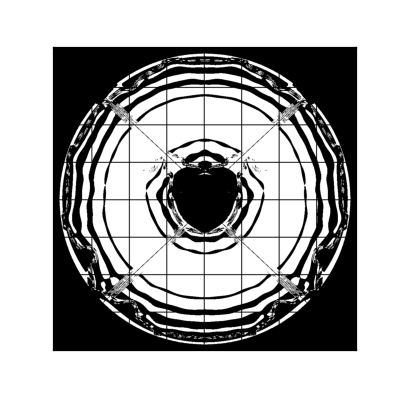}\\

&
\includegraphics[width=4cm]{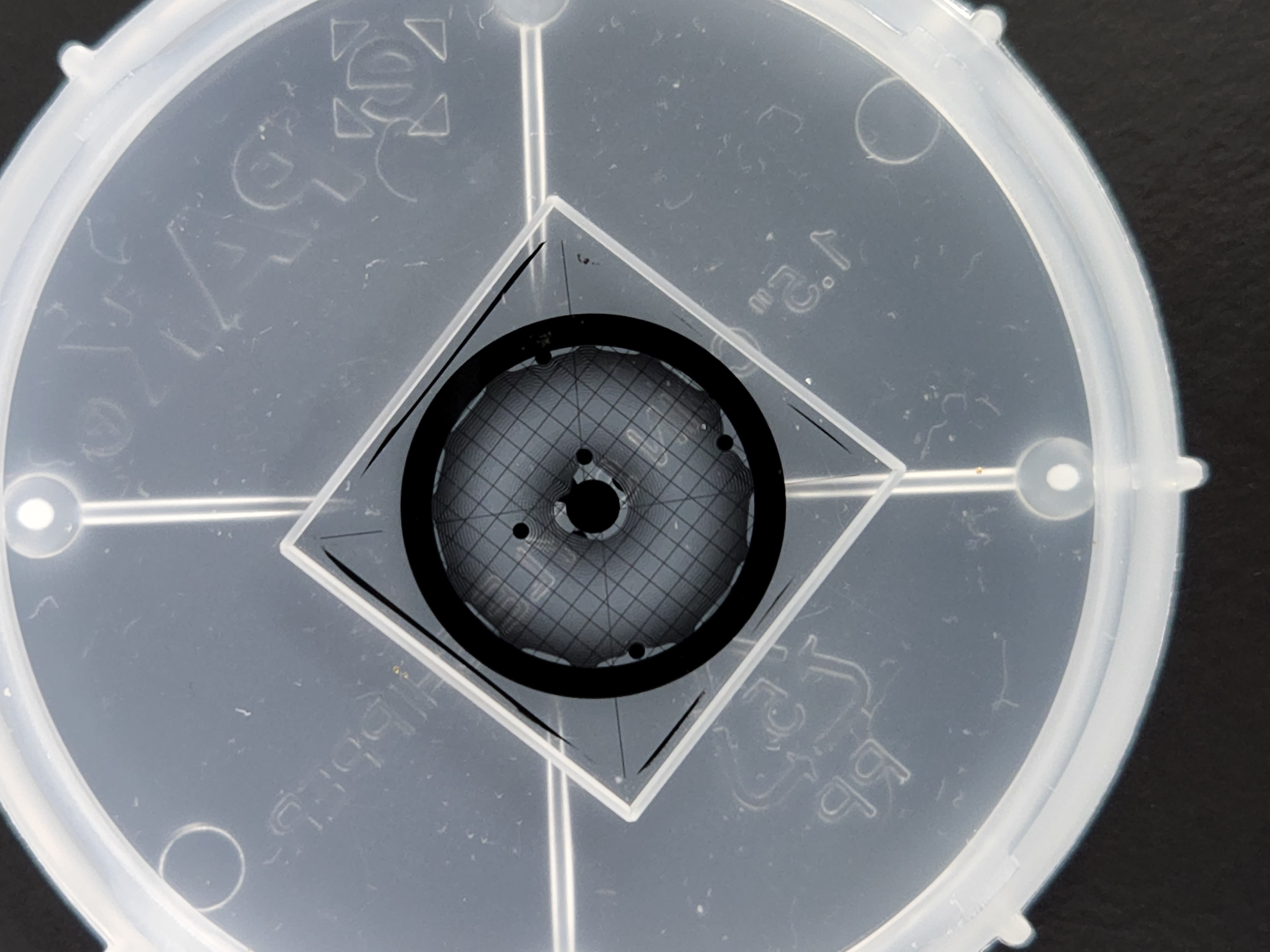} &
\includegraphics[width=4cm]{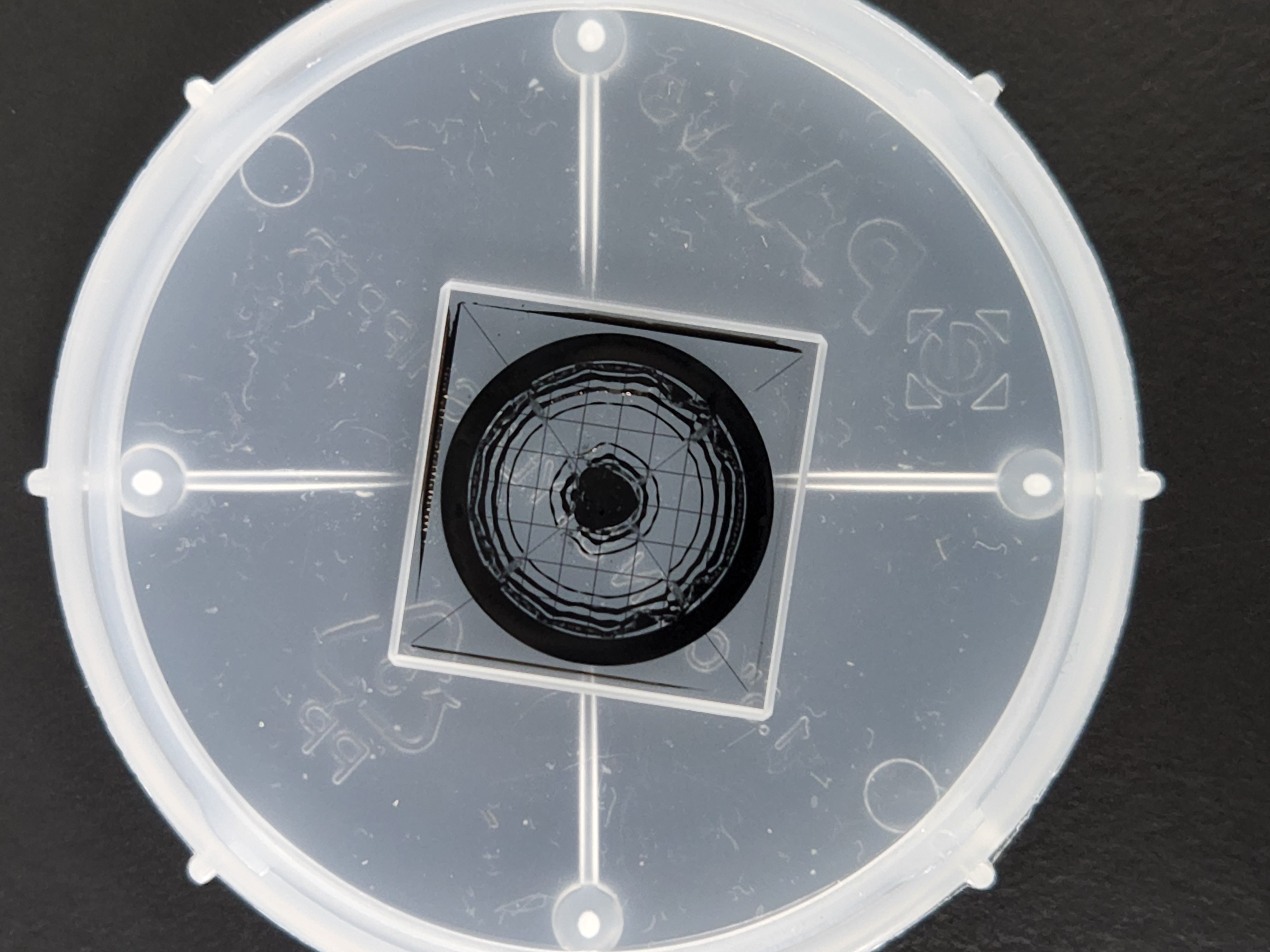}  \\

& & \\
\hline
& & \\

\textbf{LS03Symm LS} & \textbf{DualPlane LS} & \textbf{DualPlaneSymm LS}\\

\includegraphics[height=5cm]{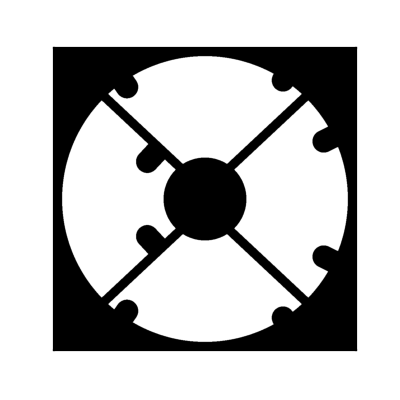} &
\includegraphics[height=5cm]{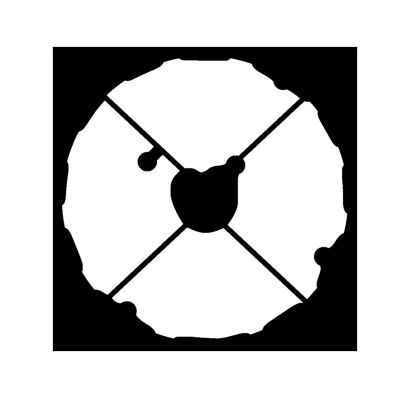} &
\includegraphics[height=5cm]{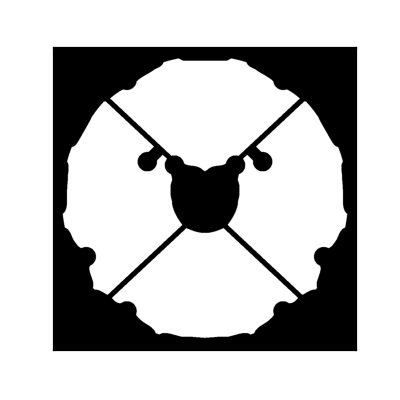}\\

&
\includegraphics[width=4cm]{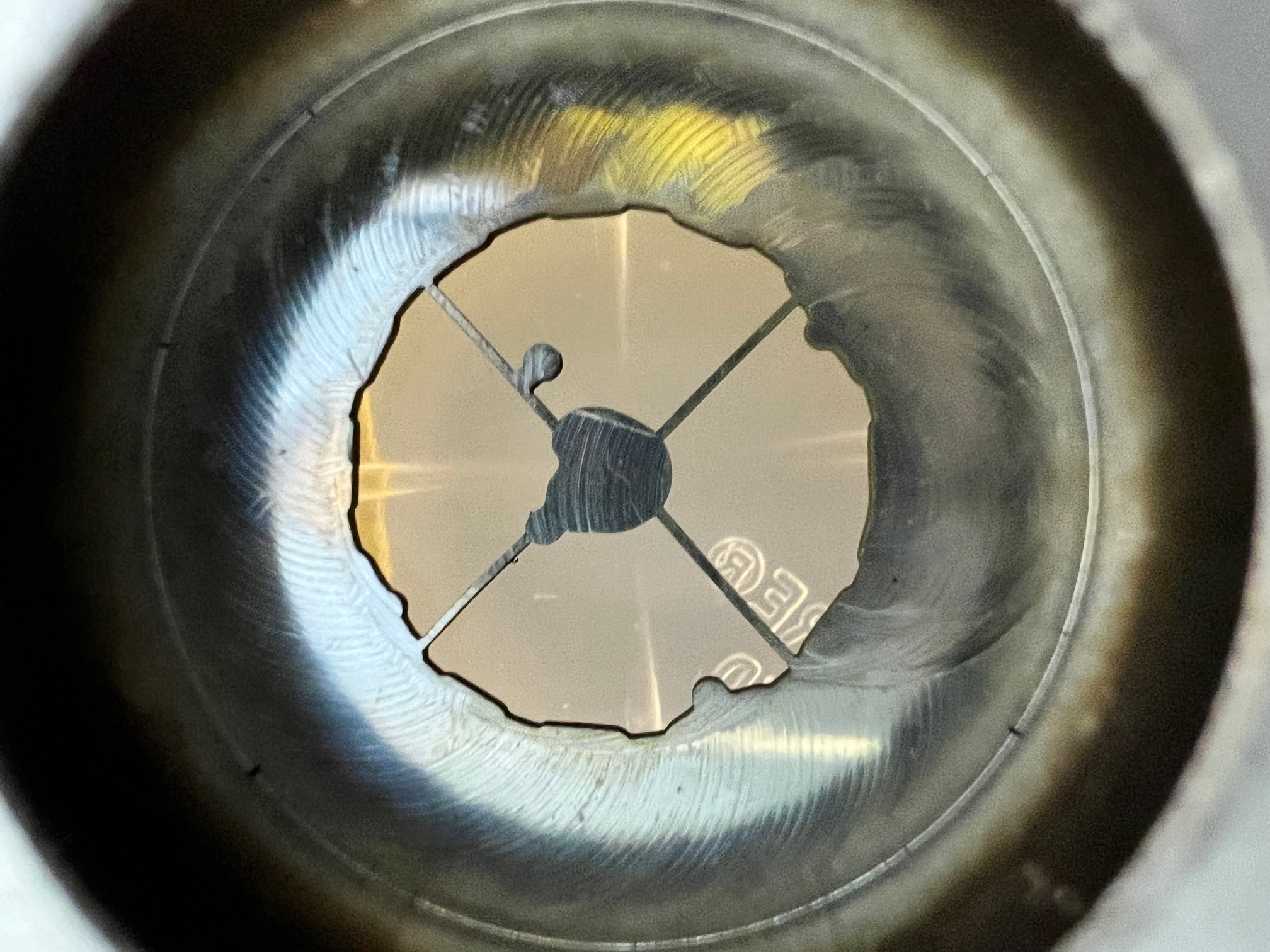} & \\
\end{tabular}
\end{center}
\caption[ap] 
{ \label{fig:apodizers} 
\textbf{Row 1)} FITS file representations of the three new apodizer designs. \textbf{Row 2)} Actual photographs taken of two of the freshly manufactured apodizers. The apodizers are inside the white square in the center of the image (the clear plastic around the outside is simply the storage case). To give a sense of scale, the diameter of each apodizer design is  approximately 1.2 cm. \textbf{Row 3)} FITS file representations of the three new LS designs (corresponding to the three new apodizer designs). \textbf{Row 4)} An actual photograph taken of one of the freshly manufactured Lyot stops. The opaque part of the LS is etched using wire-cut electrical discharge machining (EDM). To give a sense of scale, the diameter of each LS is just under 1 cm, and the struts holding the central obscuration in place have a thickness of approximately 180 microns. }
\end{figure}

\subsection{Lyot stops}
\label{sec:ls}
GPI 1.0 had nine different Lyot stop masks available on a Lyot wheel in the instrument, each having a name starting with the string "080m12" (which refers to the fact that the GPI 1.0 MEMS DM has a serial number 80 and the mask version was modified in 2012). See Footnote \footnote{http://docs.planetimager.org/pipeline/gpireference/coronagraph.html \label{footnote:gpi}} for a full description of the Lyot stop specifications. The main difference between the different Lyot stop designs was the width of the "spiders" masking the four support struts of the secondary mirror on Gemini and the width of the circular tabs masking the bad actuators (See Fig. \ref{fig:aplc_masks} for an example of a GPI 1.0 LS mask).

For the GPI 2.0 upgrade, we reduced the number of available Lyot stops from nine to just three, each corresponding to one of the three new apodizer designs described above in Section \ref{sec:ap}. This decision was made because the new apodizer designs are finely optimized to produce a target dark zone when their corresponding LS is also being used, so mix-and-matching  Lyot stops or utilizing additional Lyot stops which are not optimized for the specific apodizer in place is pointless as the resulting dark zone will produce very poor raw contrast.

\subsection{The Fabrication Process}
\label{sec:fabrication}
The original GPI 1.0 apodizers were fabricated using microdot printing as described in Sivaramakrishnan et al. (2009\cite{2009SPIE.7440E..1CS} and 2010 \cite{2010SPIE.7735E..86S}). For the GPI 2.0 upgrade, we contracted a commercial manufacturer, \textit{Advanced NanoPhotonics}\footnote{https://www.advancednanophotonics.com/}, to produce the apodizers and Lyot stops using their proprietary fabrication process\cite{2018SPIE10698E..1OS, 2022ApSS..57952250I}. The apodizer masks are transmissive, so to produce them we start with a fully transmissive substrate of infrasil glass and then pattern catalyst from which dark carbon nanotubes are grown using chemical vapor deposition at around $\sim600^{\circ}C$. The carbon nanotubes result in 100\% blocking of incoming light while minimizing reflected light (total hemispheric reflectance is 0.5\% in visible wavelengths, 0.2\% at infrared). Some of the manufacturing requirements of the apodizers are an excellent transmitted wavefront error (we achieved $\sim 5$ nm RMS) and high spectral transmission (greater than 90\% from 250-500 nm and less than 92.5\% from 500-2500 nm).

\section{THEORETICAL PERFORMANCE}
\label{sec:perf}

\begin{figure*} [ht] 
\begin{center}
\includegraphics[width=\textwidth]{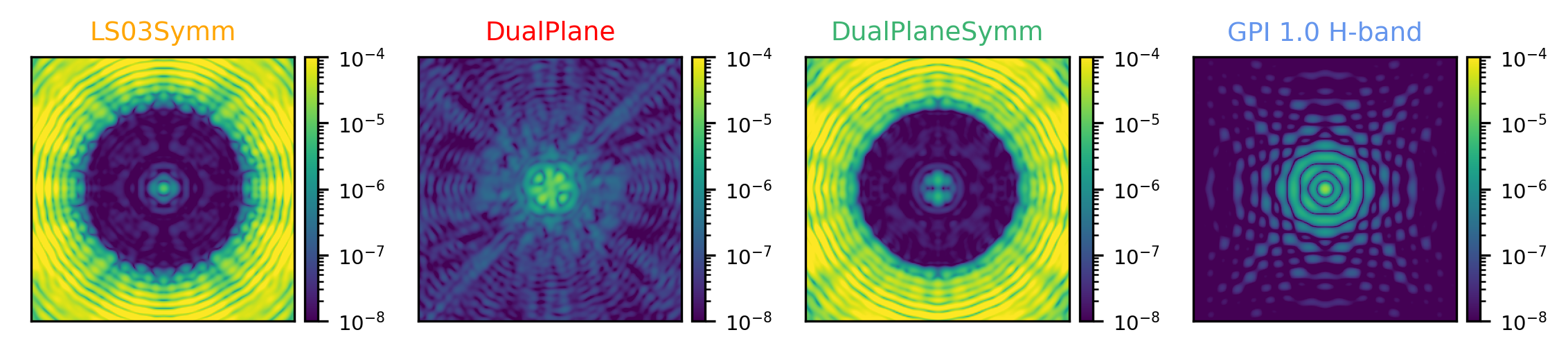}
\end{center}
\caption
{ \label{fig:coronagraphic_psfs} 
Coronagraphic PSFs (with no aberrations) for the three new designs as well as an old GPI 1.0 H-band design for reference. 
}

\begin{center}
\includegraphics[width=\textwidth]{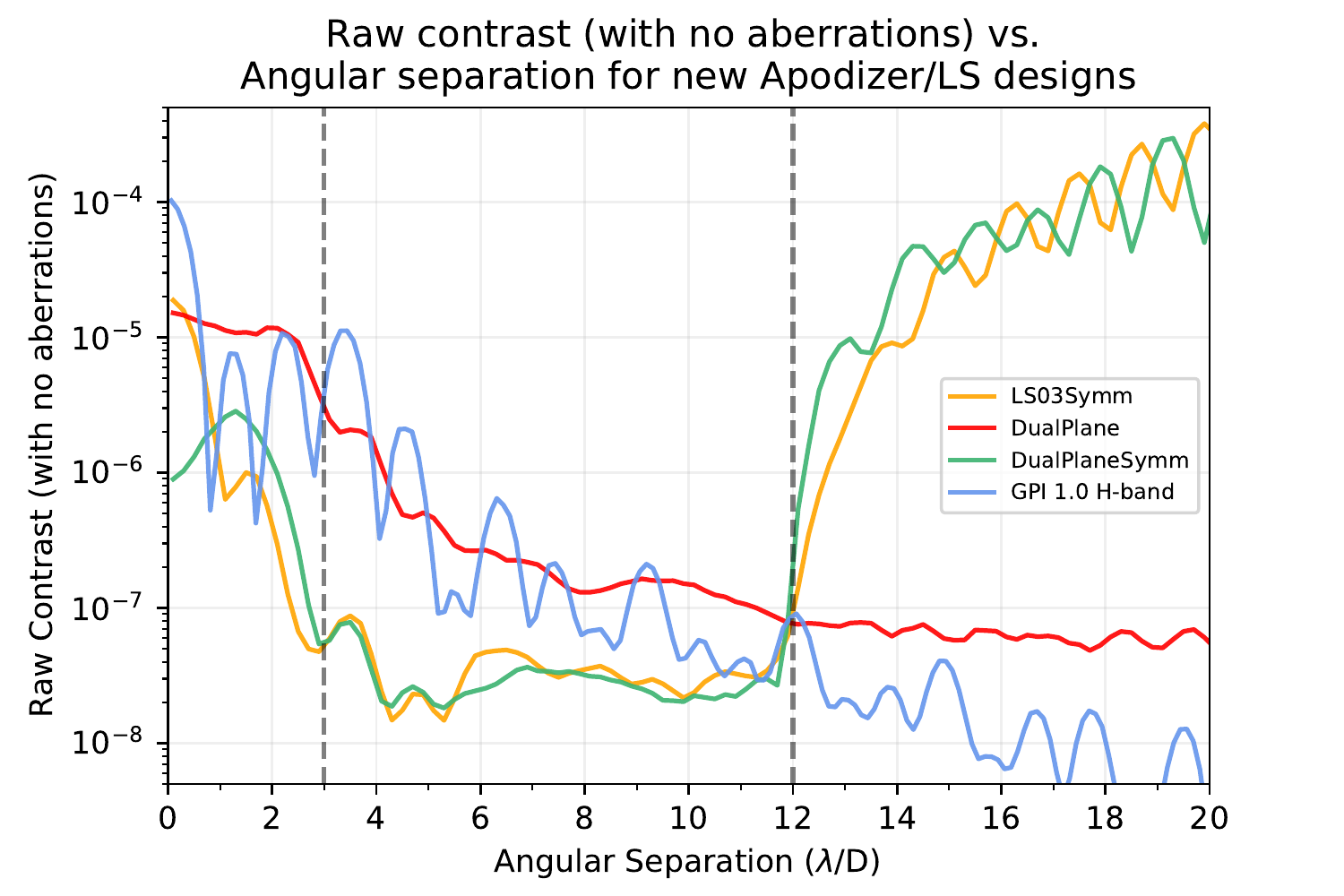}
\end{center}
\caption
{ \label{fig:raw_contrasts} 
Raw contrast (with no aberrations) vs. angular separation for the three new designs - LS03Symm (orange), DualPlane (red), DualPlaneSymm (green) - as well as an old GPI 1.0 H-band design for reference (blue). The raw contrasts are calculated directly from the coronagraphic PSFs in Fig. \ref{fig:coronagraphic_psfs}.
}
\end{figure*}

The primary metrics we use to assess the performance of a coronagraph are the raw contrast, the core throughput, the inner working angle (IWA) of the dark zone, the outer working angle (OWA) of the dark zone, and the misalignment tolerance. The first and third designs, LS03Symm and DualPlaneSymm, were optimized to have significantly better raw contrast (excluding aberrations) of approximately $10^{-8}$ across the entirety of the dark zone between the inner and outer working angle (see Figs. \ref{fig:coronagraphic_psfs} and \ref{fig:raw_contrasts}). This improved raw contrast comes at the cost of having a significantly smaller OWA compared to the GPI 1.0 masks as well as decreased throughput compared to the second design, DualPlane. The DualPlane design, has comparable raw contrast to the GPI 1.0 designs (with a slight improvement of approximately $7*10^{-6}$ raw contrast right at the IWA), but does not have the issue of a reduced OWA. In fact, the DualPlane design technically has an "infinite" OWA because the raw contrast does not go back up after some characteristic angular separation which is what characterizes an OWA in other coronagraphic dark zones. 

\begin{figure*} [ht] 
\begin{center}
\includegraphics[width=\textwidth]{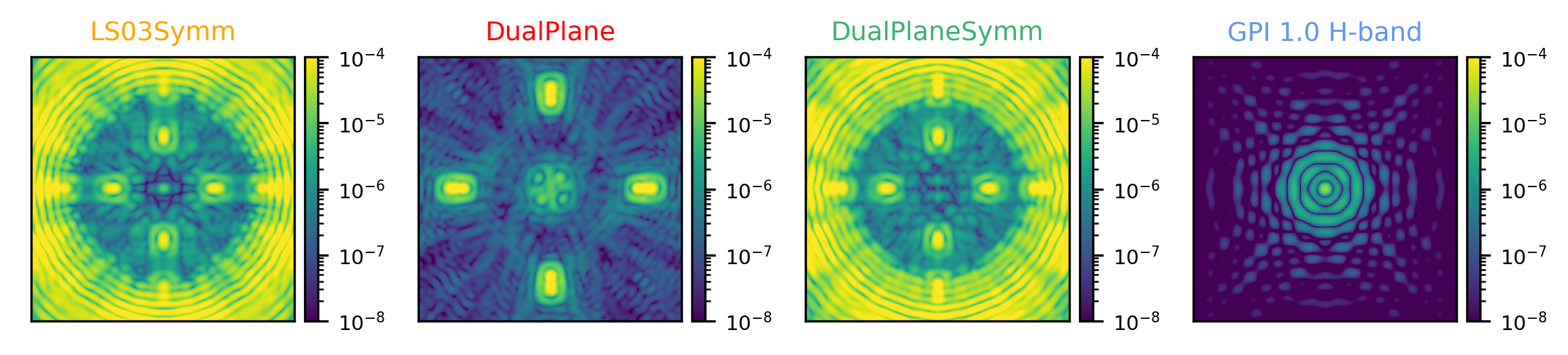}
\end{center}
\caption
{ \label{fig:coronagraphic_psfs_w_grid} 
Coronagraphic PSFs (with no aberrations) for the three new designs with the four "satspots" caused by the diffracting astrometric grid included. An old GPI 1.0 H-band design is included for reference.
}

\begin{center}
\includegraphics[width=\textwidth]{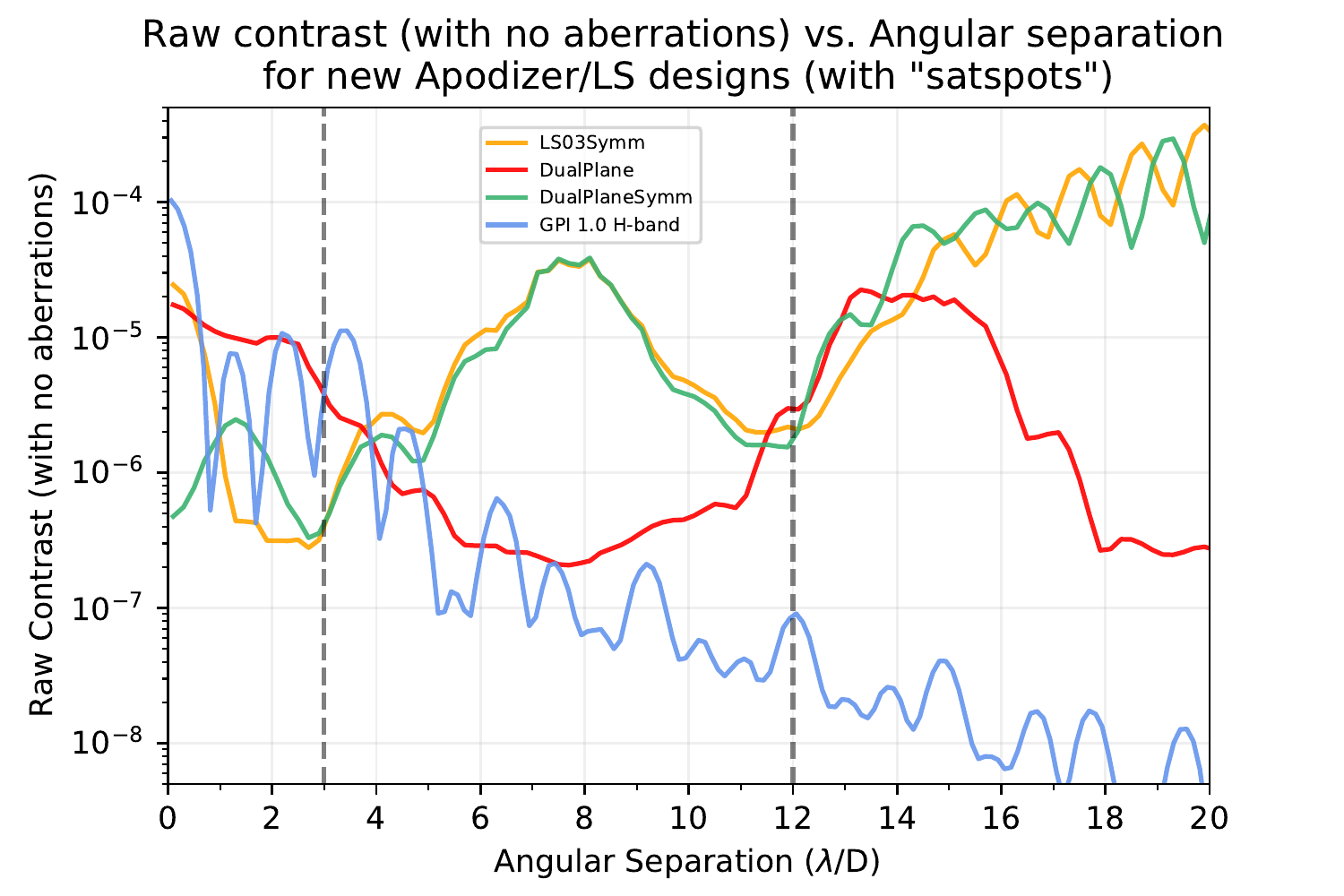}
\end{center}
\caption
{ \label{fig:raw_contrasts_w_grid} 
Raw contrast (with no aberrations) vs. angular separation for the three new designs with the four "satspots" (radially symmetric copies of the core PSF caused by the diffracting astrometric grid placed on top of the apodizers) included. Similar to Fig. \ref{fig:raw_contrasts}, we show the raw contrasts for the three new designs - LS03Symm (orange), DualPlane (red), DualPlaneSymm (green) - as well as an old GPI 1.0 H-band design for reference (blue). The raw contrasts are calculated directly from the coronagraphic PSFs in Fig. \ref{fig:coronagraphic_psfs_w_grid}.
}
\end{figure*}

The designs were optimized without the effect of the astrometric grid included. When the astrometric grid is overlaid, this produces an effect similar to a diffraction grating, causing evenly spaced copies of the core PSF to be created at four radial symmetric positions angularly separated from the central PSF - the "satspots" (short for "satellite spots"). See Figs. \ref{fig:coronagraphic_psfs_w_grid} and \ref{fig:raw_contrasts_w_grid} for the coronagraphic PSFs and raw contrast curves with the effect of the satspots included.

The nominal throughput of the new GPI 2.0 designs are either comparable or significantly improved compared to previous generation designs (see Table \ref{tab:performance_metrics}). More specifically, the first and third designs, LS03Symm and DualPlaneSymm, have comparable combined throughput (which is calculated by taking the nominal apodizer throughput and nominal LS throughput and multiplying them together) of 23.9\% and 27.1\% respectively compared to a GPI 1.0 H-band design's combined throughput of 26.0\%. The stand out case in terms of throughput is the second design, DualPlane, which has a combined throughput of 35.6\%. There is typically a trade off in coronagraph design between raw contrast and throughput, so it is natural to expect that the first and third designs perform worse in terms of throughput since they were optimized to perform significantly better in raw contrast (and vice versa for the second design). 

For a discussion of the misalignment tolerancing, see Appendix \ref{sec:LS_robustness}.

\begin{table}[ht]
\caption{Theoretical Performance Metrics}
\label{tab:performance_metrics}
\begin{center}
\begin{tabular}{|c|c|c|c|c|c|}
\hline
Design Name & Apodizer & Lyot Stop & Combined & Raw Contrast & Average Raw Contrast \\
 & Throughput & Throughput & Throughput & at (IWA) 3 $\lambda/D$ & from 3 - 12 $\lambda/D$ \\
\hline
\rule[-1ex]{0pt}{3.5ex} LS03Symm & 0.36 & 0.66 & 0.24 & $6\times10^{-8}$ & $3\times10^{-8}$\\
DualPlane & 0.49 & 0.73 & 0.36 & $3\times10^{-6}$ & $3\times10^{-7}$\\
DualPlaneSymm & 0.40 & 0.68 & 0.27 & $6\times10^{-8}$ & $3\times10^{-8}$\\
GPI 1.0 (H-band) & 0.37 & 0.70 & 0.26 & $5\times10^{-6}$ & $3\times10^{-7}$\\
\hline
\end{tabular}
\end{center}
\end{table}

\clearpage
\acknowledgments 
 
This research project was funded by the STScI Discretionary Research Fund (D0101.90238). Advanced Nanophotonics/Lambda Consulting performed research with funding from the University of Notre Dame and matching funds from the NASA SBIR program to develop transmissive apodizers for the Gemini Planet Imager upgrade. E.H.P. is supported by the NASA Hubble Fellowship grant \#HST-HF2-51467.001-A awarded by the Space Telescope Science Institute, which is operated by the Association of Universities for Research in Astronomy, Incorporated, under NASA contract NAS5-26555. The Gemini Planet Imager 2.0 project upgrade is supported by the National Science Foundation under Grant No. AST-1920180, and also significantly supported by the Heising-Simons Foundation. The international Gemini Observatory is a program of NSF’s NOIRLab, which is managed by the Association of Universities for Research in Astronomy (AURA) under a cooperative agreement with the National Science Foundation on behalf of the Gemini Observatory partnership: the National Science Foundation (United States), National Research Council (Canada), Agencia Nacional de Investigación y Desarrollo (Chile), Ministerio de Ciencia, Tecnología e Innovación (Argentina), Ministério da Ciência, Tecnologia, Inovações e Comunicações (Brazil), and Korea Astronomy and Space Science Institute (Republic of Korea).

\bibliography{report} 
\bibliographystyle{spiebib} 

\clearpage
\appendix
\section{Lyot Stop Robustness to Misalignment}
\label{sec:LS_robustness}
Based on engineering specifications and tolerances for GPI 2.0 and the Gemini North Telescope, we decided that we needed to accommodate a minimum LS misalignment tolerance of 0.5\%, which corresponds to approximately 50 microns of misalignment tolerance on the actual component (0.5\% of approximately 1cm). The LS misalignment tolerance was an important metric during the optimization process because making the LS more robust to this comes with the trade-off of reduced throughput or raw contrast. This trade-off was not well understood before we performed this study. Though we were not able to perform a full parameter space study to explore the trade-offs, we were able to run a few rounds of optimizations to explore the trade-off relationship between two of the parameters, specifically between the core throughput and the LS robustness. We ran two sets of optimizations using the Design 3 (DualPlaneSymm) template, keeping all parameters (e.g. the wavelength, FPM, OWA, target raw contrast) the same except the IWA and the LS robustness, and found there was a very clear negative, linear relationship between the core throughput and LS robustness (see Fig. \ref{fig:core_tp_vs_robustness} below). This made it fairly easy to then determine what was an optimal amount of robustness to include in the LS optimization, since we want to maximum throughput, but we need to meet the minimum LS robustness tolerance of 0.5\% which, since the designs were optimized at a resolution of 1000x1000 pixels, corresponds to a minimum tolerance of 5 pixels. 

\begin{figure*} [ht] 
\begin{center}
\includegraphics[width=\textwidth]{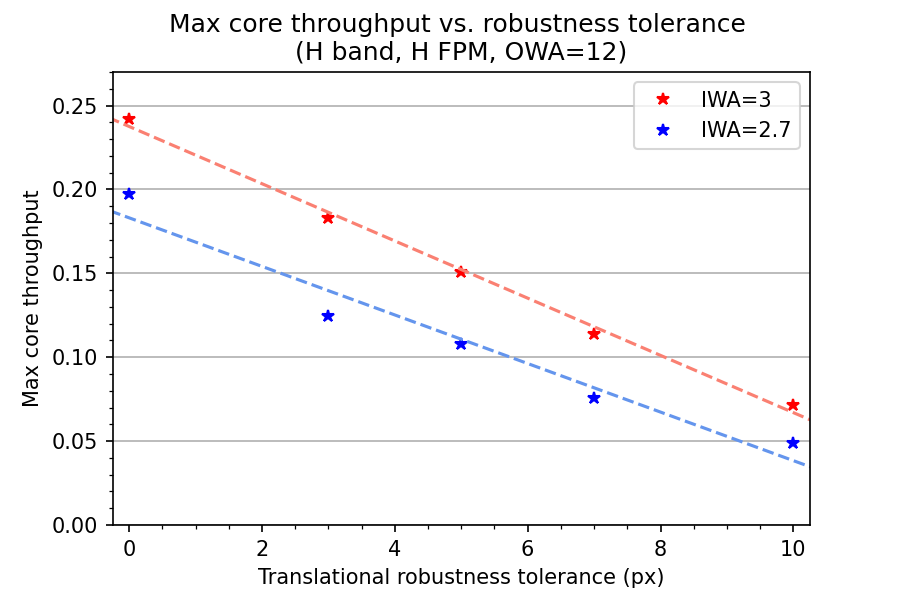}
\end{center}
\caption{ \label{fig:core_tp_vs_robustness} 
Maximum core throughput vs. LS robustness to translational misalignment for two different sets of designs (optimized for an IWA of 2.7 $\lambda$/D vs. 3 $\lambda$/D). The designs were optimized at a 1000x1000 pixel resolution, so a LS misalignment tolerance of 0.5\% corresponds to 5 pixels of translational robustness tolerance. These designs used the LS for Design 3 (DualPlaneSymm). 
}
\end{figure*} 

\end{document}